\documentstyle[pra,aps,epsfig]{revtex}
\begin{document}
\draft
\title{On the concepts of radial and angular kinetic energies}
\author{Jens Peder Dahl$^{1,2}$\footnote{Electronic address: 
jpd@kemi.dtu.dk} and Wolfgang P. Schleich$^1$}
\address{
$^1$Abteilung f\"{u}r Quantenphysik, Universit\"{a}t Ulm, 
D-89069 Ulm, Germany\\
$^2$Chemical Physics, Department of Chemistry, Technical University of 
Denmark, DTU 207, DK-2800 Lyngby, Denmark}

\date{\today}
\maketitle

\begin{abstract}
We consider a general central-field system in $D$ dimensions and show 
that the division of the kinetic energy into radial and angular parts 
proceeds differently in the wavefunction picture and the Weyl-Wigner 
phase-space picture. Thus, the radial and angular kinetic energies are 
different quantities in the two pictures, containing different 
physical information, but the relation between them is well defined. 
We discuss this relation and illustrate its nature by examples 
referring to a free-particle and to a ground-state hydrogen atom.
\end{abstract}


\section{Introduction}
 \label{intro}
Phase-space representations of quantum mechanics play an increasingly 
important role in several branches of physics, including quantum 
optics and atomic physics. The principal reason for this is the 
conceptual possibility these representations give for viewing the position 
and momentum characteristics of a quantum state in the same picture. 
Phase space is often useful for the description of stationary states, 
and it has become a natural background for describing the 
quantum-mechanical time evolution of wavepackets, for both matter 
waves and electromagnetic waves. Several phase-space representations 
have been discussed in the literature, but one of them---the so-called 
Weyl-Wigner representation---has come to play the role of a canonical 
phase-space representation, because of its simplicity 
\cite{JPD4,Wlodarz}. In accordance with this, we shall exclusively 
consider the Weyl-Wigner representation in the following.

We consider this phase-space representation to be a representation in 
its own right. In previous work \cite{JPD5,Hug,WPS} we have justified 
this statement by analyzing and solving the phase-space differential 
equations that the Wigner functions must satisfy. In particular, we 
have stressed that the Wigner functions may be determined directly 
from these equations, without reference to wavefunctions---although it 
is in general easier to determine them from the wavefunctions.

The fact that the phase-space description is a representation in its 
own right makes it relevant to apply physical intuition to the form 
and behavior of the Wigner functions, just as physical intuition may 
be applied to the form and behavior of wavefunctions. When we do this, 
we discover that our understanding of quantum states becomes enlarged, 
because the two types of intuition may work differently and therefore 
supplement each other.

In the present investigation which, for the sake of generality, is 
carried out in $D$ dimensions, we consider quantum states referred to 
a center $O$. We focus, in particular, on the evaluation of the 
angular momentum and the kinetic energy of such states. In the 
familiar picture based on wavefunctions, these quantities are 
calculated as the expectation values of operators. In the phase-space 
picture they are calculated by taking averages of dynamical 
phase-space functions with Wigner distribution functions. Performing 
the two calculations with care will, of course, lead to the same 
result. Yet, a comparison between the detailed features of the two 
descriptions leads to some interesting and physically important 
observations.

This was already noted in our previous work on the Wigner function for 
the ground state of the hydrogen atom \cite{JPD6}, in which we touched 
on a pedagogical dilemma which, for instance, has bothered writers of 
elementary textbooks \cite{Pauling}: How does one bring the fact that 
the angular momentum in the Bohr orbit is non-zero into accordance 
with the fact that the angular momentum in the Schr\"{o}dinger picture 
is zero? We referred to this dilemma as the {\em angular-momentum 
dilemma} and showed that it could be resolved by noting that the 
mapping of the operator $\hat{L}^2$ to phase space produces the 
phase-space function $|\bbox{r}\times\bbox{p}|^2 - 3\hbar^2/2$ rather 
than just $|\bbox{r}\times\bbox{p}|^2$.

In the following, we generalize this result to $D$ dimensions. In 
addition, we derive parallel but more faceted relations for 
kinetic-energy quantities, likewise in $D$ dimensions. We discuss 
these results and show that the separation of kinetic energy into a 
radial and an angular part may be done in two physically meaningful 
ways. One  is suggested by the form of the operators in the 
wave-function picture, the other by classical-like dynamical functions 
in the phase-space picture. The relation between the two variants of 
radial and angular kinetic energies is tied to the Weyl correspondence 
rule and is, therefore, well defined.

We illustrate the conceptual difference between the two types of 
kinetic-energy separation by two important examples in three 
dimensions. One is the simplest possible time-dependent state of a 
free particle, the other is the stationary ground state of the 
hydrogen atom. For the first example, we find that the phase-space 
induced separation of the kinetic energy into a radial and an angular 
part depends on time in an intuitively simple way, whereas the 
operator-based separation is independent of time. For the ground state 
of the hydrogen atom, the operator-based separation leads to an 
angular kinetic energy of zero, whereas the phase-space induced 
separation classifies the whole kinetic energy as angular kinetic 
energy. This striking difference between the results of the two types 
of separation is well reflected in the form of the wavefunction versus 
the form of the Wigner function. It illustrates in a perfect way how 
the physical richness that is hidden in the simplest state of the 
simplest atom can only be seen by looking at the state from 
different angles. It also illustrates the intricate way in which the 
roots of classical mechanics are buried in the quantum-mechanical 
soil.

The paper, which is intended to be reasonably self contained, is 
organized in the following way: In Sec.\ \ref{HC} we give a brief 
overview of hyperspherical coordinates and the central-field form of 
wavefunctions in $D$ dimensions. In Sec.\ \ref{AMKE} we define the 
angular momentum and perform the operator-based separation of the 
kinetic energy into a radial and an angular part. We express the 
result both in terms of general operators and in terms of differential 
operators. As a background for the rest of the paper we recall the 
salient aspects of the Weyl-Wigner transformation in Sec.\ \ref{WW} . 
In Sec.\ \ref{AngMom} we discuss the concept of angular momentum in 
the phase-space picture. We introduce the concepts of $q$ ({\em 
quantum}) angular momentum and $c$ ({\em classical-like}) angular 
momentum and discuss the relation between them. In Sec.\ \ref{keps} we 
give a similar discussion of the kinetic energy in phase space and of 
its separation into a radial part and an angular part. Secs.\ 
\ref{Free} and \ref{Hatom} are devoted to two illustrative examples in 
three dimensions. Sec.\ \ref{AllD} generalizes the examples to $D$ 
dimensions. Sec.\ \ref{Concl} is our conclusion.

\section{Hyperspherical Coordinates}
 \label{HC}
Let $\bbox{r} = (x_1,x_2,\ldots,x_D)$ be the position vector of a 
`particle' moving in $D$-dimensional position space \cite{note}, and 
let $\bbox{p} = (p_1,p_2,\ldots,p_D)$ be its conjugate momentum. We 
take $x_i$, and hence also $p_i$, to be Cartesian coordinates. In 
accordance with this, we introduce the hyperradius $r$ by the relation 
$r^2 = x_1^2 + x_2^2 + \ldots + x_D^2$, and likewise $p$, the 
magnitude of the momentum, by the relation $p^2 = p_1^2 + p_2^2 + 
\ldots + p_D^2$. Quantum mechanically, we adopt the position-space 
representation and write 
  \begin{equation}
\hat{\bbox{p}} = (\hat{p}_1, \hat{p}_2, \ldots, \hat{p}_D) 
= -i\hbar \left(\frac{\partial}{\partial x_1}, 
\frac{\partial}{\partial x_2}, \ldots, 
\frac{\partial}{\partial x_D}\right),
  \end{equation}
and
  \begin{equation}
\hat{p}^2 = -\hbar^2\left(\frac{\partial^2}{\partial x_1^2} 
+ \frac{\partial^2}{\partial x_2^2} + \ldots + 
\frac{\partial^2}{\partial x_D^2}\right) 
= -\hbar^2 \nabla^2 ,
  \end{equation}
where $\nabla^2$ is the $D$-dimensional Laplacian. The kinetic energy 
of a quantum-mechanical particle with mass $M$ is represented by the 
operator 
  \begin{equation}
\hat{T} = \frac{\hat{p}^2}{2M} = -\frac{\hbar^2}{2M}\nabla^2 \, .
 \label{T}
  \end{equation}
The central-field Hamiltonian
  \begin{equation}
\hat{H} = \frac{\hat{p}^2}{2M} + V(r)
  \end{equation}
determines the motion of the particle in a central field $V(r)$. 

To introduce hyperspherical coordinates in position space, one writes 
$x_i=r\eta_i$, where the $\eta_i$'s are $D$ functions of $D-1$ angular 
coordinates. Both the angles and the $\eta_i$'s may be chosen in 
different ways, but a choice similar to the following one is generally 
used, albeit with varying notation for the angles:
  \begin{eqnarray}
x_1 & = & r \sin\theta_{D-1} \sin\theta_{D-2} \ldots \sin\theta_2 
\sin\theta_1 \nonumber \\
x_2 & = & r \sin\theta_{D-1} \sin\theta_{D-2} \ldots \sin\theta_2 
\cos\theta_1 \nonumber \\
x_3 & = & r \sin\theta_{D-1} \sin\theta_{D-2} \ldots \cos\theta_2 
\nonumber \\
 & \vdots & \nonumber \\
x_{D-1} & = & r \sin\theta_{D-1} \cos\theta_{D-2} \nonumber \\
x_D & = & r \cos\theta_{D-1} 
 \label{HS}
  \end{eqnarray}
with ($0\leq r \leq \infty$), ($0 \leq \theta_1 < 2\pi$), 
($0 \leq \theta_2 < \pi$), \ldots, ($0 \leq \theta_{D-1} < \pi$). 
A similar representation may, of course, be set up for the vector 
$\bbox{p}$ in momentum space. 

For the volume element in position space we have the expression 
  \begin{equation}
d\bbox{r} = dx_1dx_2\ldots dx_D = r^{D-1}dr d\Omega, 
 \label{dv}
  \end{equation}
where the solid-angle element $d\Omega$ is given by 
  \begin{eqnarray}
d\Omega & = & (\sin\theta_{D-1})^{D-2} (\sin\theta_{D-2})^{D-3} 
\ldots\sin\theta_2 \nonumber \\
 & & \times d\theta_{D-1}\ldots d\theta_2 d\theta_1 .
  \end{eqnarray}
Integrating over all angles gives the total solid angle:
  \begin{equation}
S_D = \int d\Omega = \frac{2\pi^{D/2}}{\Gamma(D/2)} \, .
 \label{SD}
  \end{equation}

Wavefunctions of the central-field problem are conveniently referred 
to basis functions of the form
  \begin{equation}
\psi(\bbox{r}) = R(r) Y(\Omega)
  \end{equation}
where $\Omega$ is a collective notation for the angular coordinates 
$(\theta_1, \theta_2, \ldots, \theta_{D-1})$, and $Y(\Omega)$ is 
a hyperspherical harmonic. The hyperspherical harmonics were 
introduced and extensively studied by Green \cite{Green} and Hill 
\cite{Hill}. They have also been much studied by later authors. (See, 
in particular, the comprehensive presentations by Sommerfeld 
\cite{Sommerfeld}, Louck \cite{Louck} and Avery \cite{Avery}.) The 
hyperspherical harmonics are eigenfunctions of the operator 
$\hat{L}^2$ which represents the square of the total angular momentum 
and is defined below.

\section{Angular Momentum and Kinetic Energy}
 \label{AMKE}
The angular-momentum tensor in $D$ dimensions is defined by the 
operators 
  \begin{equation}
\hat{L}_{ij} = x_i \hat{p}_j - x_j \hat{p}_i, \qquad i\neq j.
  \end{equation}
The square of the total angular momentum is
  \begin{equation}
\hat{L}^2 = 
\frac{1}{2}\sum_{i=1}^D {\sum_{j=1}^D}{}^\prime \hat{L}_{ij}^2,
 \label{L2}
  \end{equation}
where the prime indicates that the double sum excludes terms for which 
$i=j$. The angular-momentum operators are independent of $r$. They 
merely depend upon the angular coordinates.

We shall now separate the kinetic-energy operator $\hat{T}$ into two 
distinctive parts. To accomplish this in a manner that eases the 
subsequent transition to the phase-space representation, we begin by 
decomposing the $D$-dimensional unit matrix $\bbox{1}$ as follows: 
  \begin{equation}
\bbox{1} = \frac{\bbox{S}}{r^2} + 
\frac{1}{2}\sum_{i=1}^D {\sum_{j=1}^D}{}^\prime 
\frac{\bbox{T}^{ij}}{r^2} \, .
 \label{resolve}
  \end{equation}
The matrices $\bbox{S}$ and $\bbox{T}^{ij}$ are defined by the 
relations 
  \begin{equation}
S_{kl} = x_k x_l
  \end{equation}
and 
  \begin{eqnarray}
(\bbox{T}^{ij})_{kl} & = & x_i^2 \delta_{kj}\delta_{lj}
+ x_j^2 \delta_{ki}\delta_{li} 
- x_ix_j(\delta_{ki}\delta_{lj} + \delta_{kj}\delta_{li}), 
\nonumber \\
 & & i\neq j, 
  \end{eqnarray}
or,
  \begin{equation}
\bbox{S} = \left(
\begin{array}{cccc}
x_1^2 & x_1x_2 & \cdots & x_1x_D \\
x_2x_1 & x_2^2 & \cdots & x_2x_D \\
\cdot & \cdot & \cdots & \cdot \\
x_Dx_1 & x_Dx_2 & \cdots & x_D^2
\end{array}
\right)
  \end{equation}
and, for instance, 
  \begin{equation}
\bbox{T}^{12} = \left(
\begin{array}{ccccc}
x_2^2 & -x_1x_2 & 0 & \cdots & 0 \\
-x_2x_1 & x_1^2 & 0 & \cdots & 0 \\
0 & 0 & 0 & \cdots & 0 \\
\cdot & \cdot & \cdot & \cdots & \cdot \\
0 & 0 & 0 & \cdots & 0 
\end{array}
\right)\, .
  \end{equation}

Adopting a dyadic notation, we may then write
  \begin{equation}
\hat{T} = \frac{1}{2M} \hat{\bbox{p}}\cdot \left(
\frac{\bbox{S}}{r^2} + 
\frac{1}{2}\sum_{i=1}^D {\sum_{j=1}^D}{}^\prime 
\frac{\bbox{T}^{ij}}{r^2}
\right) \cdot \hat{\bbox{p}} \, .
  \end{equation}
Next, we note that 
  \begin{equation}
\hat{\bbox{p}}\cdot \frac{\bbox{S}}{r^2} \cdot \hat{\bbox{p}} 
= \left(\hat{\bbox{p}}\cdot\frac{\bbox{r}}{r}\right) 
\left(\frac{\bbox{r}}{r}\cdot\hat{\bbox{p}}\right)
  \end{equation}
and
  \begin{equation}
\hat{\bbox{p}}\cdot \frac{\bbox{T}^{ij}}{r^2} \cdot \hat{\bbox{p}}
= \frac{\hat{L}_{ij}^2}{r^2} \, .
  \end{equation}
Hence, the kinetic-energy operator may be written
  \begin{eqnarray}
\hat{T} & = & \hat{T}_{rad} + \hat{T}_{ang} \nonumber \\
 & = & \frac{1}{2M} 
\left(\hat{\bbox{p}}\cdot\frac{\bbox{r}}{r}\right) 
\left(\frac{\bbox{r}}{r}\cdot\hat{\bbox{p}}\right)
+ \frac{\hat{L}^2}{2Mr^2} \, ,
 \label{Tab}
  \end{eqnarray}
where $\hat{T}_{rad}$ has the form
  \begin{equation}
\hat{T}_{rad} = \frac{1}{2M} 
\left(\hat{\bbox{p}}\cdot\frac{\bbox{r}}{r}\right) 
\left(\frac{\bbox{r}}{r}\cdot\hat{\bbox{p}}\right).
 \label{Ta}
  \end{equation}
It represents the radial kinetic energy. $\hat{T}_{ang}$, which represents 
the angular kinetic energy, is given by the operator 
  \begin{equation}
\hat{T}_{ang} = \frac{\hat{L}^2}{2Mr^2} \, .
 \label{Tb}
  \end{equation}

A modified expression for $\hat{T}_{rad}$ may be obtained by introducing 
the {\em radial momentum} $\hat{p}_r$ by the definition
  \begin{equation}
\hat{p}_r = \frac{1}{2}\left( \frac{\bbox{r}}{r}\cdot\hat{\bbox{p}}
+ \hat{\bbox{p}}\cdot\frac{\bbox{r}}{r} \right) 
 \label{pr}
  \end{equation}
and realizing that 
  \begin{equation}
\left(\hat{\bbox{p}}\cdot\frac{\bbox{r}}{r}\right) 
\left(\frac{\bbox{r}}{r}\cdot\hat{\bbox{p}}\right) 
= \hat{p}_r^2 + \frac{\hbar^2}{4r^2}(D-1)(D-3).
  \end{equation}
This gives
  \begin{equation}
\hat{T}_{rad} = \frac{\hat{p}_r^2}{2M} + 
\frac{\hbar^2 (D-1)(D-3)}{8Mr^2} \, .
 \label{Ta2}
  \end{equation}

To express $\hat{T}_{rad}$ as a differential operator, we note from 
Eq.\ (\ref{HS}) that the definition of hyperspherical coordinates 
implies that 
  \begin{equation}
\frac{\partial}{\partial r} = 
\sum_{i=1}^D \frac{\partial x_i}{\partial r}
\frac{\partial}{\partial x_i} = 
\sum_{i=1}^D \frac{x_i}{r}\frac{\partial}{\partial x_i} = 
\frac{\bbox{r}}{r}\cdot\nabla.
  \end{equation}
Using this, and the commutation relations between the components of 
$\bbox{r}$ and $\hat{\bbox{p}}$, turns the expression (\ref{Ta}) into 
the following forms 
  \begin{eqnarray}
\hat{T}_{rad} & = & -\frac{\hbar^2}{2M}\frac{1}{r^{D-1}}
\frac{\partial}{\partial r} r^{D-1}\frac{\partial}{\partial r} 
\nonumber \\
 & = & -\frac{\hbar^2}{2M}
\left(\frac{\partial^2}{\partial r^2} 
+ \frac{D-1}{r}\frac{\partial}{\partial r}\right).
  \end{eqnarray}
These are familiar expressions. 

\section{The Weyl-Wigner Transformation}
 \label{WW}
Let $\psi(\bbox{r})$ be a normalized position-space wavefunction,
   \begin{equation}
\int \psi^*(\bbox{r}) \psi(\bbox{r}) d\bbox{r} = 1.
 \label{norm}
  \end{equation}
Further, let $\hat{A}$ be some operator acting on $\psi$. The 
expectation value of $\hat{A}$ in the state $\psi$ is then
  \begin{equation}
\langle \hat{A}\rangle = 
\int \psi^*(\bbox{r})\hat{A} \psi(\bbox{r}) d\bbox{r}.
 \label{expect}
  \end{equation}
When $\hat{A}$ is Hermitian, $\langle \hat{A}\rangle$ is real.

Another way to evaluate the expectation value (\ref{expect}) is by 
introducing the $2D$-dimensional $(\bbox{r},\bbox{p})$ phase space, 
and then use the expression 
  \begin{equation}
\langle \hat{A}\rangle = 
\int\int a(\bbox{r},\bbox{p}) W(\bbox{r},\bbox{p}) 
d\bbox{r} d\bbox{p}.
  \end{equation}
Here, $W(\bbox{r},\bbox{p})$ is the Wigner function corresponding 
to the wavefunction $\psi(\bbox{r})$, and $a(\bbox{r},\bbox{p})$ 
is the dynamical phase-space function corresponding to the 
operator $\hat{A}$. For a Hermitian operator 
$a(\bbox{r},\bbox{p})$ is real. The Wigner function 
$W(\bbox{r},\bbox{p})$ is always real, but may take negative 
values. Its integral over phase space is, however, always equal to 1,
  \begin{equation}
\int\int W(\bbox{r},\bbox{p}) d\bbox{r} d\bbox{p} = 1.
  \end{equation}
It has the particle density $\rho(\bbox{r})$ and the momentum 
density $\Pi(\bbox{p})$ as marginal densities:
\begin{mathletters}
  \begin{eqnarray}
\rho(\bbox{r}) & = & \int W(\bbox{r},\bbox{p}) d\bbox{p} ,\\
\Pi(\bbox{p}) & = & \int W(\bbox{r},\bbox{p}) d\bbox{r} .
  \end{eqnarray}
\end{mathletters}

The Wigner function is defined as follows 
\cite{Wigner,Groenewold,Moyal,Hillery,JPD4,WPS}
  \begin{eqnarray}
W(\bbox{r},\bbox{p}) & = & \left(\frac{1}{2\pi\hbar}\right)^D
\int \psi^*\left(\bbox{r}-\bbox{r}^\prime/2\right)
\psi(\bbox{r}+\bbox{r}^\prime/2) \nonumber \\
 & & \times e^{-i\bbox{p}\cdot\bbox{r}^\prime/\hbar} 
d\bbox{r}^\prime .
  \end{eqnarray}
If the operator $\hat{A}$ is of the form $F(\hat{\bbox{r}}) + 
G(\hat{\bbox{p}})$, then $a(\bbox{r},\bbox{p})$ will be simply 
$F(\bbox{r})+G(\bbox{p})$. Otherwise, the noncommutativity between 
$\hat{\bbox{r}}$ and $\hat{\bbox{p}}$ will come into play. The 
transformation involved is the Weyl transformation 
\cite{Weyl,Groenewold,Moyal,Hillery,JPD4,WPS}. It may, for instance, 
be represented in the following form
  \begin{equation}
a(\bbox{r},\bbox{p}) = \int \langle\bbox{r}+\bbox{r}^\prime/2|
\hat{A}|\bbox{r}-\bbox{r}^\prime/2\rangle 
e^{-i\bbox{p}\cdot\bbox{r}^\prime/\hbar} d\bbox{r}^\prime ,
  \end{equation}
in which the matrix element of $\hat{A}$ is defined with respect 
to two eigenstates of the position-vector operator, corresponding
to the eigenvalues $\bbox{r} + \bbox{r}^\prime/2$ and 
$\bbox{r} - \bbox{r}^\prime/2$, respectively.

Let $\hat{C}=\hat{A}\hat{B}$. The dynamical phase-space function 
corresponding to the operator $\hat{C}$ is then given by the star 
product
  \begin{equation}
c(\bbox{r},\bbox{p}) 
= a(\bbox{r},\bbox{p}) * b(\bbox{r},\bbox{p}),
  \end{equation}
 where
  \begin{eqnarray}
c(\bbox{r},\bbox{p}) & = & \exp\left[\frac{i\hbar}{2}\left(
\frac{\partial}{\partial\bbox{r}_1}\cdot
\frac{\partial}{\partial\bbox{p}_2} - 
\frac{\partial}{\partial\bbox{p}_1}\cdot
\frac{\partial}{\partial\bbox{r}_2} \right)\right] \nonumber \\
 & & a(\bbox{r},\bbox{p}) b(\bbox{r},\bbox{p}).
 \label{star}
  \end{eqnarray}
Here, the subscript 1 on a differential operator indicates that this 
operator acts only on the first function in the product 
$a(\bbox{r},\bbox{p}) b(\bbox{r},\bbox{p})$. Similarly, the subscript 
2 is used with operators that only act on the second function in the 
product.


\section{Angular momentum in the Phase-Space Picture}
 \label{AngMom}

We denote the Weyl transform of the operator $\hat{L}_{ij}$ by 
$\Lambda_{ij}$. It is simply given by
  \begin{equation}
\Lambda_{ij} = x_i p_j-x_j p_i.
  \end{equation}
In analogy with Eq.\ (\ref{L2}) we introduce the dynamical 
phase-space function
  \begin{equation}
\Lambda^2 = 
\frac{1}{2}\sum_{i=1}^D {\sum_{j=1}^D}{}^\prime \Lambda_{ij}^2.
 \label{Lambda2}
  \end{equation}
This function is, however, not the Weyl transform of $\hat{L}^2$. 

To determine the actual Weyl transform of $\hat{L}^2$, we first 
determine the Weyl transform of the operator $\hat{L}_{ij}^2$. We do 
this by invoking the relation (\ref{star}), with $a$ and $b$ both 
equal to $\Lambda_{ij}$. It is found that only the three first terms 
in the expansion of the exponential operator contribute to the result. 
The Weyl transform of $\hat{L}_{ij}^2$ is thus found to be 
$\Lambda_{ij}^2 - \frac{1}{2}\hbar^2$. We express the result as the 
mapping 
  \begin{equation}
\hat{L}_{ij}^2 \longmapsto \Lambda_{ij}^2 - \frac{1}{2}\hbar^2.
  \end{equation}
From this, we get for the square of the total angular momentum:
  \begin{equation}
\hat{L}^2 \longmapsto \Lambda^2 - \frac{D(D-1)}{4}\hbar^2.
 \label{L2qc}
  \end{equation}
This is the generalisation of the result for $D=3$ that we gave 
in the Introduction.

Hence we have, for a state described by the wavefunction 
$\psi(\bbox{r})$ and the Wigner function $W(\bbox{r},\bbox{p})$, that
  \begin{eqnarray}
\langle \hat{L}^2\rangle & = & 
\int \psi^*(\bbox{r})\hat{L}^2 \psi(\bbox{r}) d\bbox{r} 
\nonumber \\
 & = & \int\int \Lambda^2 W(\bbox{r},\bbox{p}) 
d\bbox{r} d\bbox{p} - \frac{D(D-1)}{4}\hbar^2.
 \label{qcang}
  \end{eqnarray}
To put this and the other relations above into perspective, let us 
distinguish between two equally well defined angular momenta. One of 
these goes naturally with the wavefunction description. The other goes 
naturally with the phase-space description. The first angular 
momentum, which we shall call the $q$ ({\em quantum}) angular 
momentum, is defined by the operators $\hat{L}_{ij}$ and $\hat{L}^2$. 
The other angular momentum we shall call the $c$ 
({\em classical-like}) angular momentum. It is defined by the 
dynamical phase-space functions $\Lambda_{ij}$ and $\Lambda^2$. The 
two angular momenta  are connected by relations like that of Eq.\ 
(\ref{L2qc}), but the kinds of intuition one may attach to them are 
quite different.

Thus, the $q$ angular momentum is primarily an algebraic, or 
group-theoretical concept. The $\hat{L}_{ij}$ operators are generators 
of infinitesimal rotations, and their eigenvalues describe the 
possible behavior of a given state under rotations. An $s$-state in a 
three-dimensional world, with its zero-eigenvalue of $\hat{L}^2$, is 
for instance invariant under a rotation about any axis through the 
origin. This is what the eigenvalue of $\hat{L}^2$ tells us. The 
position-space wavefunction $\psi(\bbox{r})$ is independent of angles, 
and hence it predicts all directions of $\bbox{r}$ to be equally 
probable. If we prefer to describe the state by its momentum-space 
wavefunction 
  \begin{equation}
\phi(\bbox{p}) = \left(\frac{1}{2\pi\hbar}\right)^{\frac{3}{2}}
\int \psi(\bbox{r}) e^{-i\bbox{p}\cdot\bbox{r}/\hbar} d\bbox{r},
  \end{equation}
then this wavefunction is independent of angles about the origin of 
momentum space. Hence, all directions of $\bbox{p}$ are also equally 
probable. But one cannot talk about a coupling between the 
directions of $\bbox{r}$ and $\bbox{p}$ in this description. 

The expression for the $c$ angular momentum is, on the other hand, 
just the classical angular-momentum expression for particle motion 
relative to the center $O$. Thus, it provides a measure of the 
relative direction of $\bbox{r}$ and $\bbox{p}$ with respect to 
$O$. This directional correlation, which is not directly apparent from 
the wavefunction, is manifestly present in the Wigner function. It 
leads to a non-zero $c$ angular momentum even if the $q$ angular 
momentum vanishes. This is in complete accordance with the general 
relation (\ref{qcang}).

\section{Kinetic energy in the phase-space picture}
 \label{keps}
The Weyl transform of the kinetic-energy operator $\hat{T}$ given by 
the expression (\ref{T}), is simply
  \begin{equation}
T^c = \frac{p^2}{2M} \, .
  \end{equation}
To separate it into a classical-like radial part, $T_{rad}^c$, and a 
classical-like angular part, $T_{ang}^c$, we note that the resolution 
of the identity matrix given by Eq.\ (\ref{resolve}) also holds in 
phase space. Hence, we may also write 
  \begin{equation}
T^c = \frac{1}{2M} \bbox{p}\cdot \left(
\frac{\bbox{S}}{r^2} + 
\frac{1}{2}\sum_{i=1}^D {\sum_{j=1}^D}{}^\prime 
\frac{\bbox{T}^{ij}}{r^2}
\right) \cdot \bbox{p} \, .
  \end{equation}
But now the components of $\bbox{p}$ commute with the components of 
$\bbox{r}$, and the analogue of Eq.\ (\ref{Tab}) becomes 
  \begin{eqnarray}
T^c & = & T_{rad}^c + T_{ang}^c \nonumber \\
 & = & \frac{1}{2M}\left(\frac{\bbox{p}\cdot\bbox{r}}{r}\right)^2 
+ \frac{\Lambda^2}{2Mr^2}
  \end{eqnarray}
with $\Lambda^2$ given by Eq.\ (\ref{Lambda2}). The phase-space 
version of the radial kinetic energy is consequently
  \begin{eqnarray}
T_{rad}^c & = & 
\frac{1}{2M}\left(\frac{\bbox{p}\cdot\bbox{r}}{r}\right)^2
\nonumber \\
 & = & \frac{(p\cos u)^2}{2M}\, .
 \label{Tca}
  \end{eqnarray}
The last expression is obtained by putting 
$\bbox{r}\cdot\bbox{p} = rp\cos u$, where $u$ is the angle between 
$\bbox{r}$ and $\bbox{p}$.
The angular kinetic energy is
  \begin{eqnarray}
T_{ang}^c & = & \frac{\Lambda^2}{2Mr^2} \nonumber \\
 & = & \frac{(p\sin u)^2}{2M}\, .
 \label{Tcb}
  \end{eqnarray}
The validity of the last expression is simplest verified by noting 
that $p^2-(p\cos u)^2=(p\sin u)^2$. 

It is important to note that $T_{rad}^c$ is not the Weyl transform of 
$\hat{T}_{rad}$, nor is $T_{ang}^c$ the Weyl transform of 
$\hat{T}_{ang}$. To 
determine the actual Weyl transforms of $\hat{T}_{rad}$ and 
$\hat{T}_{ang}$ 
requires repeated applications of the relation (\ref{star}) to the 
expressions (\ref{Ta}) and (\ref{Tb}). We find, after some tedious 
algebra:
  \begin{equation}
\hat{T}_{rad} \longmapsto T_{rad}^c + \frac{(D-1)(D-2)\hbar^2}{8Mr^2}
 \label{mapTa}
  \end{equation}
and 
  \begin{equation}
\hat{T}_{ang} \longmapsto T_{ang}^c - \frac{(D-1)(D-2)\hbar^2}{8Mr^2} 
\, .
 \label{mapTb}
  \end{equation}

The relation (\ref{mapTa}) between the radial kinetic energies might 
also have been obtained from the relation (\ref{Ta2}) by exploiting 
the following interesting relations:
  \begin{equation}
\hat{p}_r \longmapsto \frac{\bbox{r}\cdot\bbox{p}}{r}
  \end{equation}
and 
  \begin{equation}
\hat{p}_r^2 \longmapsto 
\left(\frac{\bbox{r}\cdot\bbox{p}}{r}\right)^2 
+\frac{(D-1)\hbar^2}{4r^2} \, ,
  \end{equation}
where $\hat{p}_r$ is the operator defined by Eq.\ (\ref{pr}). These 
relations may again be derived by repeated application of the 
expression (\ref{star}) for the star product.

We shall now apply the results of this and the previous section to two 
important examples. 

\section{A free-particle state}
 \label{Free}

In this section we consider a minimum-uncertainty state in three 
dimensions. The wavefunction for such a state is 
  \begin{equation}
\psi(\bbox{r},0) = 
\left(\frac{\alpha}{\sqrt{\pi}}\right)^{\frac{3}{2}}
e^{-\frac{1}{2}\alpha^2 r^2},
 \label{psi3a0}
  \end{equation}
at a chosen initial time $t=0$. The corresponding Wigner function has 
the form 
  \begin{equation}
W(\bbox{r},\bbox{p},0) = \frac{1}{(\pi\hbar)^3}
e^{-\alpha^2 r^2 - p^2/\alpha^2\hbar^2}.
 \label{W3a0}
  \end{equation}

Let us write down the Wigner function at a later time $t$ under 
the assumption that we are dealing with a free particle. We then know 
that the Wigner function will evolve in time in exactly the same way 
as a classical phase-space distribution, that is, it will develop 
according to the classical Liouville equation \cite{Hillery,JPD4,WPS}. 
Thus, the expression for $W(\bbox{r},\bbox{p},t)$ may be obtained 
from that for $W(\bbox{r},\bbox{p},0)$ by simply replacing 
$\bbox{r}$ by $\bbox{r}-t\bbox{p}/M$. In this way we get
  \begin{equation}
W(\bbox{r},\bbox{p},t) = \frac{1}{(\pi\hbar)^3}
e^{-\alpha^2 r^2-((t/\tau)^2+1)p^2/\alpha^2\hbar^2 + 
2(t/\tau) \bbox{r}\cdot\bbox{p}/\hbar},
 \label{W3at}
  \end{equation}
where
  \begin{equation}
\tau = \frac{M}{\alpha^2\hbar}.
  \end{equation}
Since $\bbox{r}\cdot\bbox{p} = rp \cos u$, where $u$ is the angle 
between the vectors $\bbox{r}$ and $\bbox{p}$, the Wigner function 
(\ref{W3at}) only depends on the three coordinates $r$, $p$ and $u$. 
It is independent of the three Euler angles that determine the 
orientation of the $(\bbox{r},\bbox{p})$ cross in phase space. This 
independence of the Euler angles expresses the overall rotational 
invariance of the state, that is, it signifies that the $q$ angular 
momentum is zero at all times. However, evaluating the average of 
$\Lambda^2$ with the Wigner functions (\ref{W3a0}) and (\ref{W3at}) 
gives $\frac{3}{2}\hbar^2$, as it should according to the relation 
(\ref{qcang}). 

We note that all angles between $\bbox{r}$ and $\bbox{p}$ are equally 
probable at $t=0$, whereas small values of $u$ are favored for large 
values of $t$. Thus, the correlation between $\bbox{r}$ and $\bbox{p}$ 
depends on $t$. Such a statement would be impossible to defend in the 
wavefunction picture. However, in the phase-space picture it is very 
meaningful. For at $t=0$, the Wigner function (\ref{W3at}) is 
concentrated about the origin of phase space. But in the course of 
time, phase space points with non-zero values of $p$ will move to 
points with larger values of $r$, with $\bbox{r}$ and $\bbox{p}$ 
becoming more and more parallel. But this just amounts to small values 
of $u$ being favored for large values of $t$, as in the expression 
(\ref{W3at}).

This behavior is also reflected in the expectation values of the 
radial and angular parts of the kinetic energy as a function of time, 
but only in the phase-space picture. It is readily verified that the 
total kinetic energy associated with the wavefunction (\ref{psi3a0}) 
has the value 
  \begin{equation}
\langle \hat{T} \rangle = \frac{3}{4}\frac{\alpha^2\hbar^2}{M} \,.
  \end{equation}
It is, of course, independent of $t$. And since $\psi(\bbox{r},t)$ 
represents an $s$ state, the expectation value of the $\hat{T}_{ang}$ 
of Eq.\ (\ref{Tb}) is zero at all times. Thus we have:
  \begin{eqnarray}
\langle \hat{T}_{rad} \rangle & = & 
\frac{3}{4}\frac{\alpha^2\hbar^2}{M} \, , \nonumber \\
\langle \hat{T}_{ang} \rangle & = & 0,
 \label{Tcab}
  \end{eqnarray}
for all $t$.

In the phase-space picture, we must average the dynamical phase-space 
functions $T^c$, $T_{rad}^c$ and $T_{ang}^c$ with the phase-space 
function(\ref{W3at}). We denote the resulting energies by ${\cal T}$, 
${\cal T}_{rad}$ and ${\cal T}_{ang}$, respectively:
  \begin{equation}
{\cal T} = \frac{1}{2M}\int\int p^2 W(\bbox{r},\bbox{p}) 
d\bbox{r} d\bbox{p} ,
 \label{T3}
  \end{equation}
  \begin{equation}
{\cal T}_{rad} = \frac{1}{2M}\int\int (p \cos u)^2 
W(\bbox{r},\bbox{p}) 
d\bbox{r} d\bbox{p} ,
 \label{T3rad}
  \end{equation}
and
  \begin{equation}
{\cal T}_{ang} = \frac{1}{2M}\int\int (p \sin u)^2 W(\bbox{r},\bbox{p}) 
d\bbox{r} d\bbox{p} ,
 \label{T3ang}
  \end{equation}
We have, of course, that
  \begin{equation}
{\cal T} = \langle \hat{T} \rangle .
  \end{equation}

Rather than evaluating the expressions (\ref{T3rad}) and (\ref{T3ang}) 
by direct integration, we may proceed in the following analytical way: 
According to Eqs.\ (\ref{mapTa}) and (\ref{mapTb}), with $D=3$, 
$T_{rad}^c$ and $T_{ang}^c$ are the Weyl transforms of the operators 
  \begin{equation}
\hat{T}_{rad}^\prime = \hat{T}_{rad} - \frac{\hbar^2}{4Mr^2}
 \label{Taa}
  \end{equation}
and
  \begin{equation}
\hat{T}_{rad}^\prime  = \hat{T}_{ang} + \frac{\hbar^2}{4Mr^2} \, ,
 \label{Tbb}
  \end{equation}
respectively. Hence, we merely have to modify the values in 
(\ref{Tcab}) with the expectation value of $\hbar^2/4Mr^2$ to get the 
values of ${\cal T}_{rad}$ and ${\cal T}_{ang}$. Evaluating this value 
by averaging $\hbar^2/4Mr^2$ with the Wigner function (\ref{W3at}) 
gives 
  \begin{equation}
\left\langle \frac{\hbar^2}{4Mr^2} \right\rangle = 
\frac{\alpha^2\hbar^2}{2M} \frac{1}{(t/\tau)^2+1} \, .
  \end{equation}
Hence, we get
  \begin{eqnarray}
{\cal T}_{rad}  & = & \frac{\alpha^2\hbar^2}{M} 
\left(\frac{3}{4} - \frac{1}{2((t/\tau)^2+1)}\right), \nonumber \\
{\cal T}_{ang}  & = & \frac{\alpha^2\hbar^2}{M} 
\frac{1}{2((t/\tau)^2+1)} \, .
 \label{calTab}
  \end{eqnarray}
Thus, the classical-like kinetic energy is purely radial for very 
large values of $t$. However, at $t=0$ only one third of the kinetic 
energy is radial, corresponding to the value 
$(1/4)(\alpha^2\hbar^2/M)$. The angular kinetic energy is twice as 
large. This reflects the fact that $\bbox{r}$ and $\bbox{p}$ are 
entirely uncorrelated at $t=0$, so that $\bbox{p}$ is twice as likely 
to be perpendicular to $\bbox{r}$ as being parallel to $\bbox{r}$.

\begin{figure}
\begin{center}
\epsfig{width=3.5in,file=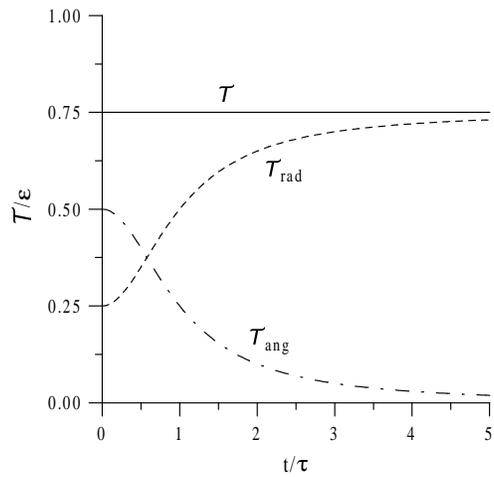}
\caption{Time dependence of the classical-like kinetic energy of a 
free particle described by the Wigner function (\ref{W3at}). $t$ is 
measured in units of $\tau = M/\alpha^2 \hbar$, energies in units of 
$\epsilon = \alpha^2\hbar^2/M$. The kinetic energy $\cal T$ is 
separated into its radial part ${\cal T}_{rad}$ and its angular part 
${\cal T}_{ang}$.} 
\label{fig1}
\end{center}
\end{figure}

The energies ${\cal T}$, ${\cal T}_{rad}$ and ${\cal T}_{ang}$ are 
shown graphically in Fig.\ \ref{fig1}, as functions of $t$.

\section{The Hydrogen Atom}
 \label{Hatom}

Having studied a free-particle case, we shall next consider a 
bound-state case, namely, the ground state of the three-dimensional 
hydrogen atom. The wavefunction is now 
  \begin{equation}
\psi(\bbox{r}) = \sqrt{\frac{1}{\pi a_0^3}} e^{-r/a_0}\, ,
 \label{psiH}
  \end{equation}
where $a_0$ is the Bohr radius,
  \begin{equation}
a_0 = \frac{\hbar^2}{M}\frac{4\pi\epsilon_0}{e^2}.
  \end{equation}
$M$ is the electron mass (we treat the nucleus as being infinitely 
heavy), and $e$ is the magnitude of the elementary charge. The state 
considered is a stationary state, and the Wigner function is 
accordingly independent of time. As in the previous example, we are 
dealing with an $s$ state, so the Wigner function is again 
independent of the Euler angles that determine the orientation of the 
$(\bbox{r},\bbox{p})$ cross, but it does depend on the angle $u$ 
between $\bbox{r}$ and $\bbox{p}$, and in fact in a more complicated 
manner than in the expression (\ref{W3at}). In addition, the Wigner 
function for the hydrogen atom takes both positive and negative 
values, whereas the Wigner function (\ref{W3at}) is non-negative at 
all times.

We have made a detailed study of the Wigner function for the 
hydrogen atom in \cite{JPD6}. Our results were, {\em inter alia}, 
presented as a series of contour maps for different $u$ values. 
These maps showed that the Wigner function is everywhere positive in 
what we called the dominant subspace, that is, the part of phase space 
in which $\bbox{r}$ and $\bbox{p}$ are perpendicular ($u=\pi/2$). It 
is, in particular, large in the part of this subspace obtained by 
putting $r = a_0$ and $p = \hbar/a_0$. This is the region of phase 
space to which the ground-state motion was restricted in early quantum 
mechanics \cite{JPDbook}, since a Bohr orbit (in position space) is 
just a circle with radius $a_0$, in which the electron is supposed to 
move with the constant momentum $\hbar/a_0$. For other angles than 
$u=\pi/2$, the Wigner function develops negative regions. It is, in 
particular, strongly oscillating around the value zero for small 
angles and angles approaching $\pi$.

We shall now see that this pronounced correlation between the 
directions of $\bbox{r}$ and $\bbox{p}$ is strongly reflected in the 
partitioning of the kinetic energy. The kinetic energy associated with 
the wavefunction (\ref{psiH}) has the value
  \begin{equation}
\langle \hat{T} \rangle = \frac{\hbar^2}{2Ma_0^2} \, ,
  \end{equation}
and since $\psi(\bbox{r})$ represents an $s$ state we have, in analogy 
with the expressions (\ref{Tcab}):
  \begin{eqnarray}
\langle \hat{T}_{rad} \rangle & = & \frac{\hbar^2}{2Ma_0^2} \, ,
\nonumber \\
\langle \hat{T}_{ang} \rangle & = & 0.
  \end{eqnarray}
The relations (\ref{Taa}) and (\ref{Tbb}) still hold, and it is 
readily found that
  \begin{equation}
\left\langle \frac{\hbar^2}{4Mr^2} \right\rangle = 
\frac{\hbar^2}{2Ma_0^2} \, .
  \end{equation}
The analogue of (\ref{calTab}) becomes therefore
  \begin{eqnarray}
{\cal T}_{rad} & = & 0 , \nonumber \\
{\cal T}_{ang} & = & \frac{\hbar^2}{2Ma_0^2} \, .
  \end{eqnarray}

This is a remarkable result. For it shows that, in the phase-space 
representation, the radial kinetic energy vanishes. The kinetic energy 
is purely angular. This is of course in complete harmony with a 
picture in which the electron primarily revolves around the nucleus 
rather than moving in the radial direction.

The cases studied in this section and the previous one have both been 
for $s$ states. The expressions we have derived in the first six  
sections are, however, valid for any state independent of its angular 
momentum, but the interesting effects are most pronounced for the $s$ 
states. With a minor exception, the derived expressions are also valid 
for any dimension $D$. A few comments concerning dimensions different 
from three are, therefore, in order.

\section{Arbitrary dimensions}
 \label{AllD}
The minor exception mentioned above has to do with 
zero-angular-momentum states for $D=2$. The relations (\ref{mapTa}) 
and (\ref{mapTb}) suggest that the separation of the kinetic energy 
into a radial part and an angular part is invariant under the Weyl 
transformation for $D=2$. This is, however, not quite true. One must 
be aware that taking expectation values with the expressions 
(\ref{mapTa}) and (\ref{mapTb}) involves taking expectation values of 
$1/r^2$, and such expectation values are undefined for wavefunctions 
that stay finite at $r=0$ in a two-dimensional world. This is because 
the volume element (\ref{dv}) only contains $r$ to the first power for 
$D=2$. Hence, one cannot exploit operator relations like those of 
Eqs.\ (\ref{Taa}) and (\ref{Tbb}) for zero-angular-momentum  states. 
This, however, does not reduce the significance of the dynamical 
phase-space functions $T_{rad}^c$ and $T_{ang}^c$. It just implies 
that the radial and angular kinetic energies ${\cal T}_{rad}$ and 
${\cal T}_{ang}$ associated with them must be evaluated directly by 
averaging with the Wigner function, using the two-dimensional version 
of the expressions (\ref{T3rad}) and (\ref{T3ang}). For the 
two-dimensional equivalent of the Wigner function (\ref{W3at}) this 
produces a $t$ dependence of the kinetic energies similar to that in 
Fig.\ \ref{fig1}, with the difference that the kinetic energy is 
equally distributed on its radial and angular components at $t=0$. 
This difference was to be expected since there is only one 
perpendicular direction to $\bbox{r}$ in two dimensions.

Concerning the hydrogen atom, it is interesting to note that the above 
conclusions for the ground state of the hydrogen atom in three 
dimensions remain valid for other values of $D$. Thus, the kinetic 
energy is purely radial in the wavefunction picture, but purely 
angular in the phase-space picture. The ground-state wavefunction for 
the $D$ dimensional hydrogen atom has the general form
  \begin{equation}
\psi(\bbox{r}) = 
\sqrt{\frac{1}{S_D}\frac{1}{a_0^D (n_0/2)^{2n_0+1}(2n_0)!}}\,
e^{-r/n_0a_0}
  \end{equation}
where $S_D$ is the total solid angle (\ref{SD}) and 
  \begin{equation}
n_0 = \frac{D-1}{2} \, .
  \end{equation}
The ground-state kinetic energy is 
  \begin{equation}
\langle \hat{T} \rangle = \frac{\hbar^2}{2Ma_0^2n_0^2} \, ,
  \end{equation}
and for $D>3$ exactly the same value is found for the expectation 
value of the $D$ dependent term in (\ref{mapTa}) and (\ref{mapTb}). 
This confirms that the classical-like kinetic energy is, in fact, 
purely angular for $D>3$.

For $D=2$, we can again not draw on expressions like those of Eqs.\ 
(\ref{Taa}) and (\ref{Tbb}). We must perform the phase-space 
integrations directly. Doing so shows that also for $D=2$, the 
classical-like kinetic energy is purely angular.

\begin{figure}
\begin{center}
\epsfig{width=3.5in,file=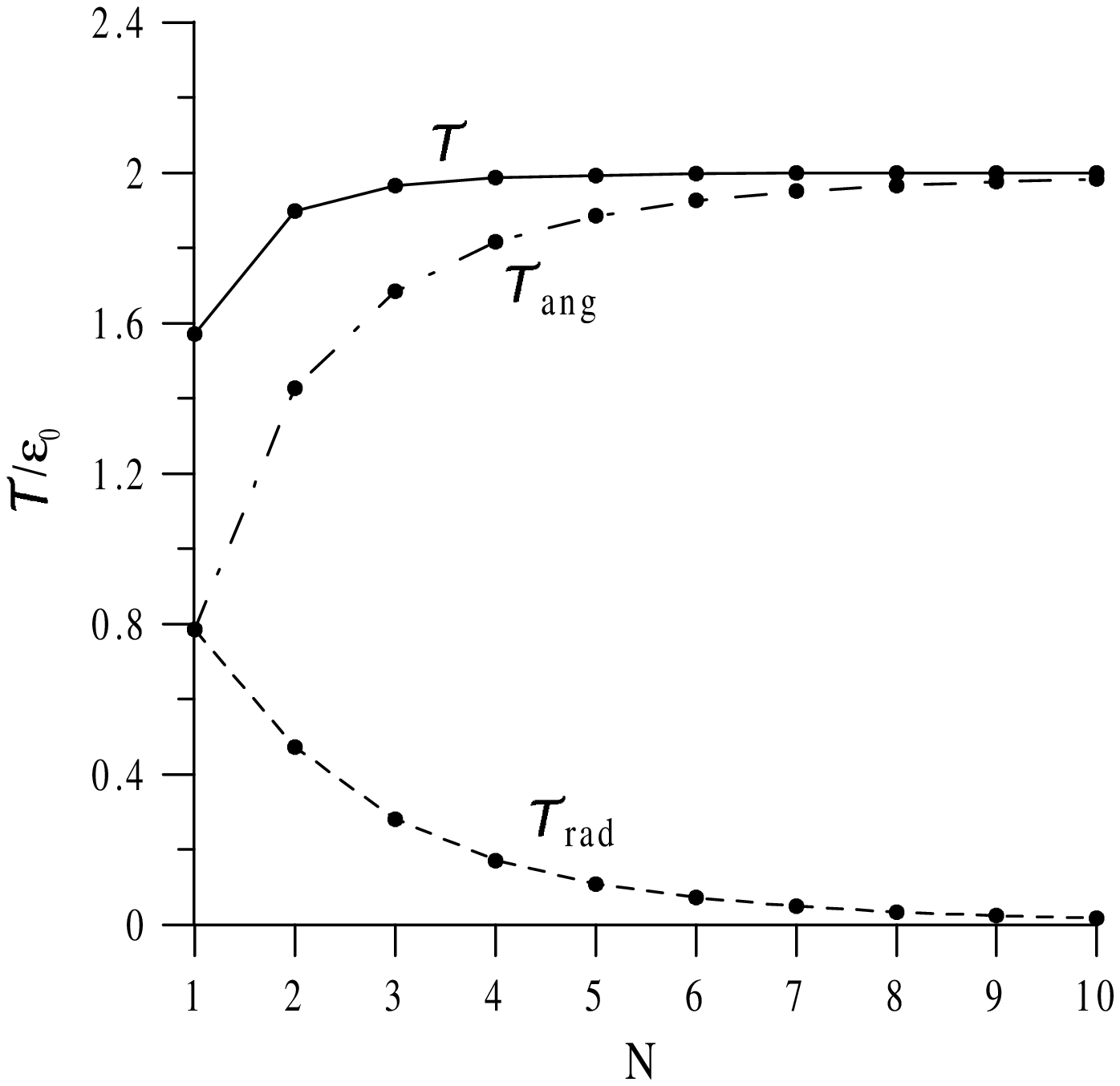}
\caption{With the ground-state wavefunction of the two-dimensional 
hydrogen atom approximated by a linear combination of $N$ Gaussians, 
the figure shows the $N$-dependence of the calculated classical-like 
kinetic energy ${\cal T}$, as well as its radial and angular parts, 
${\cal T}_{rad}$ and ${\cal T}_{ang}$. Energies are measured in units 
of $\epsilon_0 = \hbar^2/Ma_0^2$.}
 \label{fig2}
 \end{center}
\end{figure}

The said integrations over phase space are far from simple to perform, 
because the Wigner function for the hydrogen atom cannot be evaluated 
analytically \cite{JPD6,JPDdim}. A practical procedure is to expand 
the $1s$ wavefunction on a set of Gaussians. For a linear combination 
of $N$ Gaussians, the Wigner function may be determined 
analytically. The values of ${\cal T}_{rad}$ and ${\cal T}_{ang}$ may 
then be calculated by combining analytical and numerical integrations. 
For $N=1$, we have a Wigner function similar to that of Eq.\ 
(\ref{W3a0}) for a free particle, but in two dimensions only, leading 
to ${\cal T}_{rad} = {\cal T}_{ang}$. As $N$ increases, the 
contribution from ${\cal T}_{rad}$ is found to decrease, converging to 
zero for large values of $N$. This is shown graphically in 
Fig.\ \ref{fig2}. 

The information in Fig.\ \ref{fig2} is not merely numerical. It serves 
as yet another demonstration of the physical uniqueness of the Coulomb 
potential. For instead of considering the used wavefunctions to be 
approximate solutions for the Coulomb potential, we may consider them 
to be exact solutions for a different potential. As $N$ 
increases, this potential becomes more and more Coulomb like. This 
suggests that only for the Coulomb potential is the classical-like 
kinetic energy purely angular.

The linear combinations of Gaussians used to prepare Fig.\ \ref{fig2} 
were determined by the variational principle, following the 
prescription given in reference \cite{JPDdim}. In that work, 
we made explicit studies of the Wigner function for the $D$ 
dimensional hydrogen atom and presented contour curves for selected 
values of $D$. These contour maps show, {\em inter alia}, that the 
oscillations between negative and positive values of the Wigner 
function become weaker and weaker for higher $D$ values. We may 
therefore say that the phase-space distributions become more classical 
as the dimensionality $D$ increase.

\section{Discussion}
 \label{Concl}
The problem of separating the kinetic energy of a particle moving in a 
central field into a radial part and an angular part has not attracted 
much attention in the past. It has been tacitly assumed that the 
separation offered by the wavefunction picture was the only sensible 
one. In the present paper we have challenged this view by putting 
focus upon the separation offered by the phase-space picture. This 
separation is equally well defined. In contrast to the former, it 
throws much light on the correlation between the directions of the 
position vector $\bbox{r}$ and the momentum vector $\bbox{p}$. There 
is no way of discussing this correlation in the wavefunction picture.

The most amazing result of our analysis is probably our finding that 
the kinetic energy in the ground state of the hydrogen atom is purely 
radial in the wavefunction picture, and purely angular in the 
phase-space picture. This finding, which relates to arbitrary 
dimensions, reflects the great difference in the kind of intuition 
one may apply within the two pictures. This difference is also well 
illustrated by focusing upon the angular momentum. In the wavefunction 
picture, the angular momentum primarily refers to the behavior of the 
state under rotations. In the phase-space picture it also refers to 
the correlation between the directions of $\bbox{r}$ and $\bbox{p}$. 

To properly understand a quantum system one must look at it from 
different angles. The present paper shows once again that the 
phase-space picture is a fruitful supplement to the wavefunction 
picture.

\section*{Acknowledgments}
J. P. Dahl gratefully acknowledges the support of the Danish Natural 
Science Research Council, the Alexander von Humboldt Foundation, and 
the great hospitality enjoyed at the Abteilung f\"{u}r 
Quantenphysik. The work of W. P. Schleich is partially supported by 
DFG. J. J. W\l odarz is thanked for his comments on the manuscript.



\begin{thebibliography}{99}
\bibitem{note}We use the word particle in a generalized sense. The $D$ 
Cartesian coordinates $(x_1,x_2,\ldots,x_D)$ could, for instance, 
stand for the total set of Cartesian coordinates of $n$ identical 
particles, instead of the $D$ coordinates of a single individual 
particle.
\end{thebibliography}
\end{document}